\input epsf
\epsfverbosetrue

\documentstyle{l-aa}

\begin{document}
\newcommand{\ana}{A\&A~}  
\newcommand{\apj}{{ApJ~}}     
\newcommand{\aj}{{AJ~}}     
\newcommand{\mnras}{{MNRAS~}} 
\newcommand{\ea}{{et~al.~}}
\newcommand{\cgs}{$10^{-10}$ W cm$^{-2}$ $\mu$m$^{-1}$ sterad$^{-1}$}
\newcommand{\ang}{\AA \ }
\thesaurus{ 09.04.1 11.09.1 11.09.4 11.19.2 13.09.1 13.09.4
}
\title{Detection of widely distributed UIR band emission in the disk of NGC~891
\thanks{Based on observations made with ISO, an ESA project 
with instruments
funded by ESA member states (especially the PI countries: France, Germany,
the Netherlands and the United Kingdom) and with the
participation of ISAS and NASA}}
\author{K. Mattila\inst{1}
 \and K. Lehtinen\inst{1}
 \and D. Lemke\inst{2}
}
\institute{Observatory, P.O. Box 14, FIN-00014 University of Helsinki, Finland
(email: mattila@cc.helsinki.fi)
 \and Max-Planck-Institut f\"ur Astronomie, K\"onigstuhl 17, D-69117
Heidelberg, Germany 
}

\date{Received 7 July 1998; accepted date}
\offprints{K. Mattila, mattila@cc.helsinki.fi}

\maketitle

\begin{abstract}

The spectrum  of the unidentified infrared (UIR) emission bands between 
5.9 and 11.7 $\mu$m has
been observed for the first time in the disk of 
an external galaxy.
We have used the
low-resolution spectrometer of the ISOPHOT instrument aboard ISO.
The UIR bands at 6.2, 7.7, 8.6, and 11.3 $\mu$m have absolute
intensities which are 
similar to the values observed for the diffuse emission
of our own Galaxy. The UIR bands between 5.9 and 11.7 $\mu$m contribute 
$\sim$9\% of the total IR radiation of NGC~891.  
The intensity ratios and band widths in the NGC~891 disk emission are similar
to the diffuse emission of the Milky Way, pointing to a common 
carrier for the UIR bands in the two galaxies. However, there are some 
notable variations of the band ratios along the major axis of the galaxy.

\end{abstract}

\section{Introduction}   

The family of the so-called unidentified infrared emission bands (UIR bands),
centered at 3.3, 6.2, 7.7, 8.6, and 11.3 $\mu$m, has been observed
for more than 20 years (Gillett et al. 1973) in a number of
galactic objects, such as planetary nebulae (PN), HII regions, and reflection
nebulae (RN) around early type stars, as well as in the nuclear regions of 
some external galaxies. 
These objects are characterized by UV radiation densities several orders of
magnitude above the average interstellar radiation field in the galactic 
disks and as a consequence the UIR band intensities are high. 

A surprisingly high level of mid-IR emission from the disk of our Galaxy 
was first detected by Price (1981) and later confirmed by the 12 and 25 $\mu$m
IRAS observations.
Similarly,
the global IR emission of other spiral galaxies has been observed to
have a strong excess in the IRAS 12 and 25 $\mu$m bands.
This emission constitutes ca. 20 \% of the 
IR radiation energy from the disk of our Galaxy, and $\sim$ 10 -- 30 \%
of the total IR flux of other spiral galaxies (cf. Helou et al. 1991). 
Puget et al. (1985)
have proposed that the 12 $\mu$m emission of our Galaxy is largely due to the
UIR bands. This proposal has been supported by the photometric detection
of the 3.3 $\mu$m and 6.2 $\mu$m UIR bands in the galactic disk emission 
using the AROME balloon-borne experiment (Giard et al. 1988; Ristorcelli
et al. 1994) and has been confirmed by the ISO and IRTS spectrophotometry
(Mattila et al. 1996; Onaka et al. 1996). 

Strong UIR band emission at 3.3, 7.7, 8.6, and 11.3 $\mu$m has been
detected in HII region type galactic nuclei showing vigorous star formation 
activity (Moorwood 1986, Roche et al. 1991). The average
ratio of the 8.6 and 11.3 $\mu$m band intensities has been found to be higher 
in galaxy nuclei than in the nebular sources in our Galaxy (Aitken \&
Roche 1984).
With the Infrared Space Observatory (ISO) large
surveys and several detailed individual studies have been performed 
of the UIR bands in the nuclear or starburst regions of galaxies (see e.g.
Spoon et al. 1998, Moorwood et al. 1996, Boulade et al. 1996).

In the case of disks  of external 
galaxies no spectra have been obtained for the UIR bands so far.
However, in analogy to the ISM in our Galaxy, model calculations have been
presented which suggest that the UIR bands and/or continuum emission by
other transiently heated very small grains do dominate the IRAS 12 $\mu$m 
emission in other galactic disks as well ( Walterbos \& Schwering 1987, 
Xu \& DeZotti 1989, Helou et al. 1991, Walterbos \& Greenwalt 1996).

The present investigation was initiated to answer the following questions:
(1) Are the UIR emission bands at 6.2, 7.7, 8.6, and 11.3 $\mu$m
present also in the disks of external galaxies where the UV ISRF is
typically 100 -- 1000 times lower than in the galactic nuclear regions.
 (2) Can the IRAS 12 $\mu$m emission of galaxy disks be explained
(solely) as the integrated emission of the UIR bands?
 (3) What are the relative intensities of the UIR bands in the disks in comparison
with the nuclear regions of galaxies and the bright galactic sources, the PNs, 
RNs, and the HII regions?
 (4) What is the spatial distribution of the UIR band carriers in the
 disks of galaxies?
 By finding the answers we hope to contribute also to the problem of
identifying the UIR band carriers.

In this paper we present results of an ISOPHOT low resolution 
spectrometer (PHT-S) observing project
aimed at studying the UIR bands in the galaxy NGC~891.
NGC~891 is an Sb spiral closely similar to our own Galaxy, 
thus motivating a comparison
with spectrophotometry of the diffuse galactic UIR emission
(Mattila et al. 1996). NGC~891 is seen almost exactly edge-on (inclination
$89^{\circ}$). This is an essential condition in order to have a sufficiently
large column density of ISM along the line of sight to enable 
spectrophotometric 
analysis with PHT-S. For the distance of NGC~891 we adopt the value 9.5 Mpc
(van der Kruit \& Searle 1981). It is consistent with the value 8.9 Mpc
estimated by Handa et al. (1992) on the basis of the infrared Tully-Fisher
relation. 

\section{Observations and reductions}  
\label{obsreduc}

The observations were carried out during the
ISO revolutions 656 ( 1 September 1997) and 788 ( 11 January 1998). 
A linear raster of 20 positions along the major axis of NGC~891 at position 
angle  $24^{\circ}$ was observed with  PHT-S  (see Lemke et al.1996). 
The integration
time per raster position was 512 sec along the northern and 356 sec along the
southern part of the major axis. 
The raster step size was 24", i.e. equal to the PHT-S aperture size 
(24" $\times$ 24").
The observed positions with their orientation
are shown in Fig. 2 superimposed on an optical image of the galaxy.
At the distance of NGC~891 the aperture size corresponds
to 1.1 kpc. The outermost raster positions were at 216'' (9.9 kpc) North
and at -240'' (11.0 kpc) South of the central position ($\alpha$ = 2h~19m~24.6s,
$\delta$ = 42$^{\circ}$~07'~21'' (1950.0)).
Four off-axis positions were observed in addition, displaced by 24'' 
perpendicular to the major axis from the central and the 120'' North
positions. An off-position observation 5' West of the galaxy centre 
was performed in revolution 788 but it turned out to be of lower quality
than the rest of the measurements. Thus, the four off-axis positions plus
the outermost positions at the ends of the major axis were used for the 
background subtraction.

\epsfxsize=8.8cm
\begin{figure} [ht]
     \vspace{0.0cm}
     \epsfbox{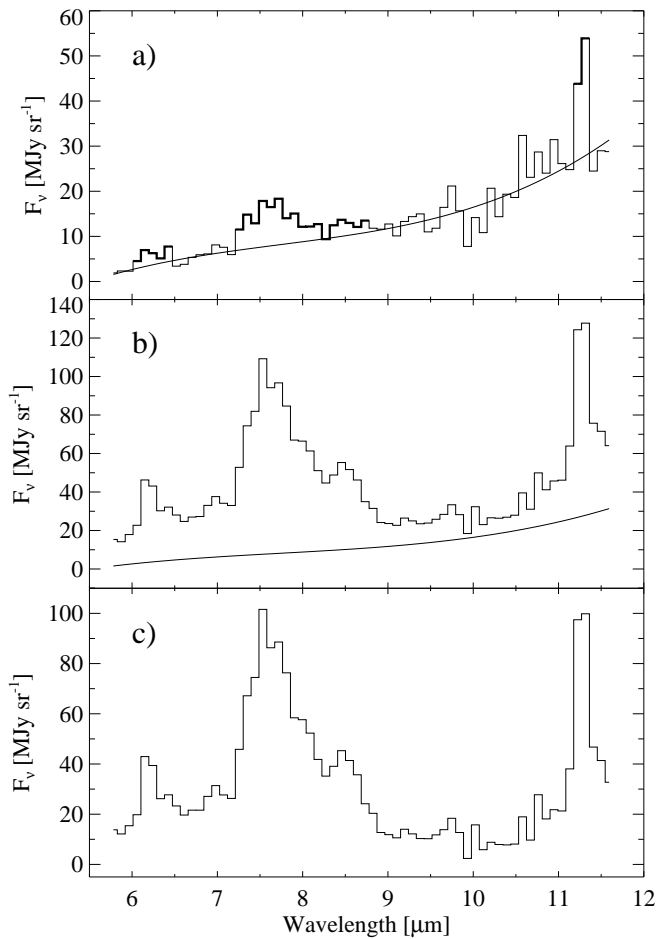}
     \vspace{0.0cm}
     \caption{{\bf a-c.} Reduction steps for PHT-S spectra. 
{\bf a} The averaged background emission spectrum 
(zodiacal emission plus possible instrumental effects) 
as observed outside the galaxy disk with the fitted 3rd 
order polynomial. {\bf b} Observed spectrum at the central position of 
NGC~891 and the polynomial fit estimate for the background. {\bf c.} Spectrum
at central position with background subtracted}
     \label{spectra}
\end{figure}

\epsfxsize=12cm
\begin{figure*} [ht]
     \epsfbox{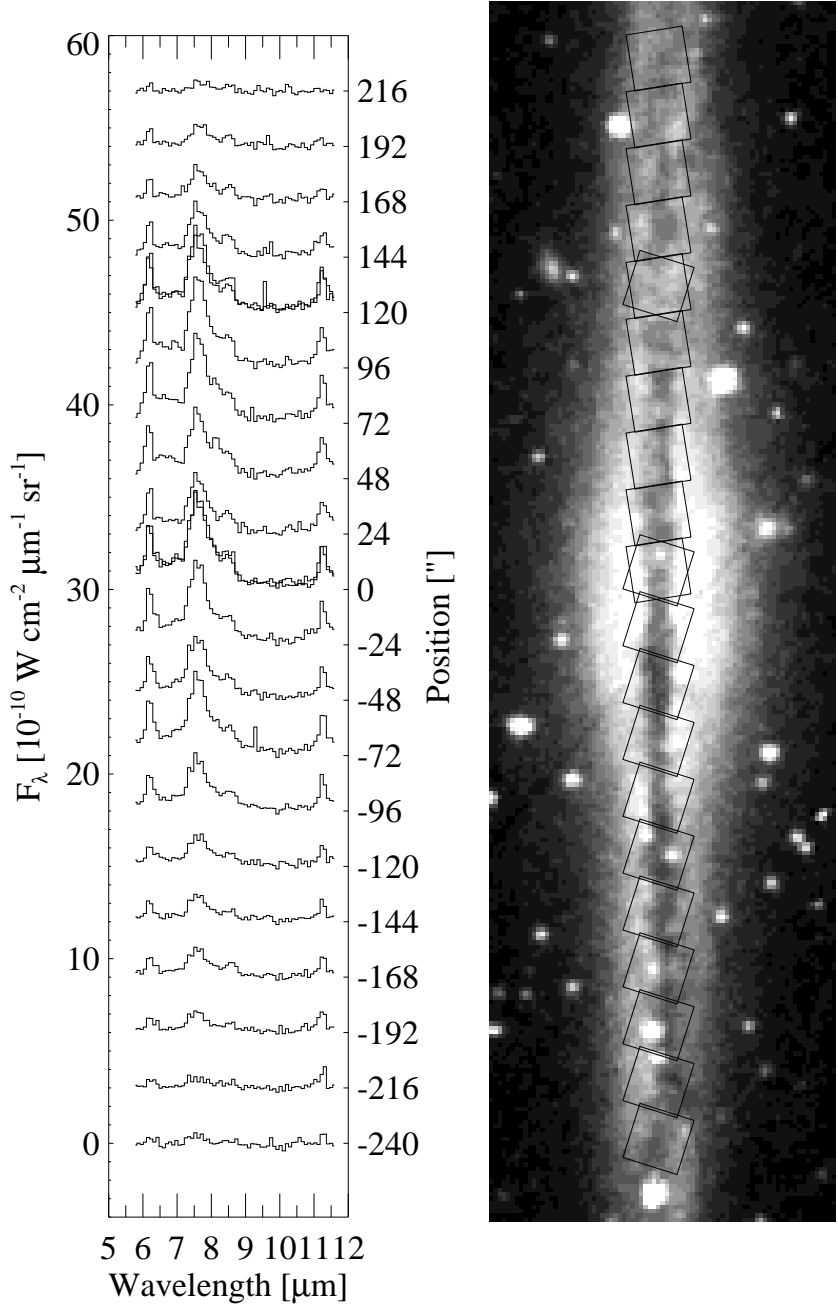}
     \caption{ {\bf a and b. a (on the left) }PHT-SL spectra along the major axis of NGC~891. 
{\bf b (on the right) } Optical B band image of NGC~891 from the Digital Sky Survey
with the observed 24''$\times$24'' areas superposed. For the central and the 120''~N
positions two spectra and two different aperture orientations corresponding to
ISO orbits 656 and 788 are shown }
     \label{photo}
\end{figure*}

The data reduction was performed using the ISOPHOT Interactive Analysis Program
(PIA) Version 7.0 (Gabriel et al. 1998).
The following reduction steps were
applied:\\
(1) Deglitching of the data (to eliminate cosmic ray effects)\\
(2) Deletion of first $\sim$ 10 sec of integration (to eliminate detector 
drift effects). \\
(3) Subtraction of dark current. Instead of the short (32 sec) dark current 
measurement preceeding each PHT-S spectrum measurement we used the 
``default'' dark values provided by PIA7.0 in dependence of ISO's orbital
position. They are based on extensive compilation of dark signal measurements
randomly distributed over the duration of the orbital science window 
(see Klaas et al. 1998, \'Abrah\'am et al. 1998).\\
(4) Calibration to convert instrumental units (V/s) to surface brightness
units (MJy sr$^{-1}$). The calibration for the PHT-S measurements 
is based on observations of the ISO standard stars HR~6705, HR~6688, HR~6817, 
and HR~6514. For the calibration of extended sources that fill the aperture 
(as in our case) the
measured beam profile for each spectrometer detector
pixel  has been integrated over the whole PHT-S aperture
and the resulting correction factors are included in the PIA7.0 calibration
values (see Klaas et al. 1998).\\
(5) Subtraction of the zodiacal emission for the long-wavelength (=PHT-SL) 
spectra (5.9 -- 11.7 $\mu$m). We used the off-axis spectra
and the outermost on-axis positions at the northern and southern end of 
the major axis to determine the foreground zodiacal emission. 
Because these positions still
contain some UIR band emission from the galaxy we excluded the wavelength
ranges of the bands and fitted a 2nd or 3rd order polynomial through the
continuum points. Some continuum emission from the galaxy
might be present in these background positions making our zodiacal emission
estimate too high. However, if we assume that the band-to-continuum ratio
for the background positions is the same as for the strong on-axis emission 
this error in the continuum correction is negligible. 
We illustrate the zodiacal background subtraction for revolution 788 in Fig. 1.
Panel (a) shows the average of the southernmost (-240'') on-axis and the two
off-axis spectra at +120'' North. The three spectra were in good agreement
with each other. UIR band emission at 6 -- 6.5, 7.3 -- 8.8, and 11 - 11.5
$\mu$m appears to be present. These portions, indicated by thick line,
were excluded and a 3rd order polynomial was fitted through the remaining
channels. The derived background spectrum is consistent with
the spectral shape and absolute level of the zodiacal emission at the
target coordinates $\lambda - \lambda_{\odot} =  102^{\circ}$ to 
$137^{\circ}$, 
$\beta = 26.5^{\circ}$  (see \'Abrah\'am et al. 1998, 
Reach et al. 1996). We point out that no spectral features at the UIR
or other wavelengths are seen in the very deep zodiacal emission spectra
presented by \'Abrah\'am et al. (1998) and Leinert et al. (1998).
The zodiacal emission spectrum has a pure continuum in this wavelength
region. The weak UIR band emission seen in our OFF positions (Fig.~1a)
is thus not connected to the zodiacal foreground emission.
Panel (b) shows  the observed spectrum for
central position of NGC~891 together with the fitted background spectrum.
After subtraction of the zodiacal light the final
spectrum of the galaxy's radiation is obtained and is
shown in panel (c).

(6) Subtraction of the zodiacal light (and any remaining instrumental effects)
for the short-wavelength(= PHT-SS) spectra (2.5 -- 4.9 $\mu$m). 
This was performed
using spectra at the same background positions in the outskirts of NGC~891 
as in the long-wavelength case. In this case, however, no polynomial fit
was done but the averaged background spectrum was used as such.

The resulting PHT-SL spectra  
 along the major axis of NGC~891 are shown in Fig.~2a. The surface
brightness scale is valid for the bottom-most spectrum at -240'' and 
the consequtive spectra are displaced by 3 units in the vertical direction.
In Fig.~2b we show an optical image (blue light) of NGC~891 reproduced
from the Digital Sky Survey. The scaling of Fig.~2b is such that 24'' 
correspond to the spacing between the spectra in Fig.~2a. 
The observed PHT-S 24''x24''aperture positions are shown in Fig.~2b with 
their correct orientations on sky, i.e. with the sides parallel to the 
spacecraft Y and Z axes.

The central and the 120'' North positions were observed twice,
once in each of the revolutions 656 and 788. The two independent measurements
agreed within 10~\%  for the PHT-SL spectra (see Fig. 2a), the rev. 788 
intensities being systematically smaller than the rev. 656 ones by this amount.
We brought the PHT-SL measurements to a uniform scale by applying a 
multiplicative correction factor of 1.05 and 0.95, to rev. 656 and 788, 
respectively. 
 The resulting surface brightness values are expected
to have an absolute accuracy of better than 30 \% (Klaas et al. 1998a, 1998b).
However, since the observations were performed as two interconnected
one-dimensional raster maps the {\em relative point-to-point} accuracy
is better than the absolute one. We estimate it to be $\sim$ 10\%. 
The relative feature intensities within the spectrometer band are 
estimated to have an uncertainty of $\sim$ 10\%.


\section{Results}  
\subsection{PHT-SL spectra (5.9 -- 11.7 $\mu$m)}        

By inspecting the spectra in Fig.~2a (see also Fig.~1c) we can immediately 
list the following qualitative results:

(1) The main UIR bands at 6.2, 7.7, 8.6, and 11.3 $\mu$m
are clearly present in the disk of NGC~891 at least to a distance of 192'',
corresponding to 8.8 kpc,  both North and South of the centre. 
The spectra look very similar to those observed for the diffuse emission
of the disk of our Galaxy as well as for bright galactic objects,
such as reflection nebulae and HII regions.\\
(2) There is an apparent broad underlying plateau-like emission between 6 and 9 $\mu$m.
This is in accordance with the findings for e.g. the Orion Bar (Bregman
et al. 1989) as well as for the diffuse emission of our Galaxy 
(Mattila et al. 1996). However, with Cauchy profile fits the plateau
emission can be explained by the line wings (see Sect. 3.1).\\ 
(3) The maximum emission at 7.7 $\mu$m occurs at the central and the
96'' North positions and is 
$\sim$ 5.0 10$^{-10}$ W cm$^{-2} \mu$m$^{-1}$ sterad$^{-1}$.
This is comparable to the level of the diffuse emission in the inner disk 
of our Galaxy, $\sim$ 3.2 10$^{-10}$ W cm$^{-2} \mu$m$^{-1}$ sterad$^{-1}$,
observed with the same instrument (see Mattila et al. 1996).\\ 
(4) The continuum emission at 9.5 -- 10.5 $\mu$m is detected and its level is 
$\sim$ 5 to 10 \% of the UIR band peaks.

Some further, more quantitative results are:\\
(5) In order to compare the integrated emission of the UIR bands 
with the IRAS 12 $\mu$m emission we have summed up the PHT-S spectra 
between 7.5 and 11.6 $\mu$m 
(the IRAS band pass covers  $\sim$ 7.5 $\mu$m to 15 $\mu$m). 
The results for the total and the UIR band emission are \newline
$7.4~10^{-17}$ and $6.3~10^{-17}$ W~cm$^{-2}$, \\
respectively, integrated over all the 
observed positions. The total PHT-S flux value can be compared with  
IRAS 12 $\mu$m band flux of NGC~891, 5.66 Jy (Rice et al. 1988) which corresponds
to an in-band flux of 7.64~10$^{-17}$ W cm$^{-2}$. Taking into account
an absolute calibration error of $\sim$20\% for IRAS and $\sim$30\% for 
PHT-S    
we conclude that the two integrated flux values are in good agreement and that 
at the very least $\sim 50 - 60$ \% of  
the IRAS 12 $\mu$m emission is due to the UIR bands. 
If the relative intensities of the UIR bands in the 11.5 to 16 $\mu$m range
are similar to the highly illuminated regions in our Galaxy 
it is possible that the {\em whole}
IRAS 12$\mu$m emission of NGC 891 can be explained in terms of the UIR bands
plus the weak underlying continuum seen in our spectra.\\
(6) Gu\'elin et al. (1993) have estimated the IR flux density of the large
grains in NGC~891. They have fitted the IRAS 60 and 100 $\mu$m and their
1.3 mm flux densities with a warm ($T_{\rm wd}$ = 30 K) and a cold
($T_{\rm cd}$ = 15 K) dust component. They find the corresponding luminosities
of $L_{\rm wd} = 1.5~10^{10}$ L$_{\odot}$ and 
$L_{\rm cd} = 0.5~10^{10}$ L$_{\odot}$ which 
correspond to a flux density of 
7.2~10$^{-16}$ W cm$^{-2}$.
The energy emitted in the UIR bands at 6.2, 7.7, 8.6, and 11.3 $\mu$m 
is 1.1$\pm$0.3~10$^{-16}$ W cm$^{-2}$ or $\sim$ 15\% relative to the large grains. 
This is somewhat higher than the fraction (10 \%)
found for the diffuse emission of our Galaxy (Mattila et al. 1996).
Adding the contribution of the IRAS 12 and 25 $\mu$m bands (see Rice et al.
1988) we find that the total IR flux density of NGC~891 is 
11.9~10$^{-16}$ W cm$^{-2}$, and the UIR band contribution 
is $\sim$9\% of the total.

\epsfxsize=8.8cm
\begin{figure} [ht]
     \vspace{0cm}
     \epsfbox{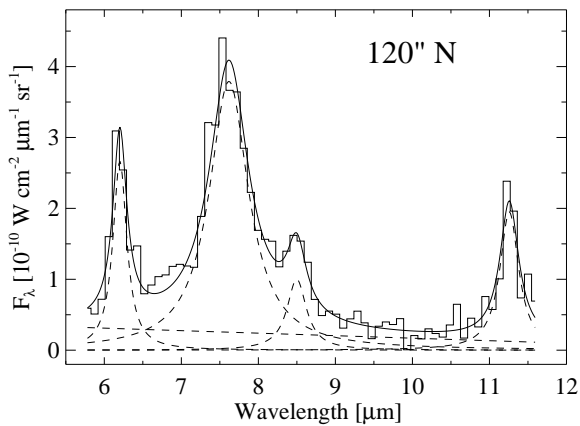}
     \vspace{0cm}
     \caption{Spectrum at 120'' N fitted with four Cauchy line profiles
plus a linear baseline } 
     \label{spectra}
\end{figure}

\epsfxsize=8.8cm
\begin{figure} [ht]
     \vspace{0.0cm}
     \epsfbox{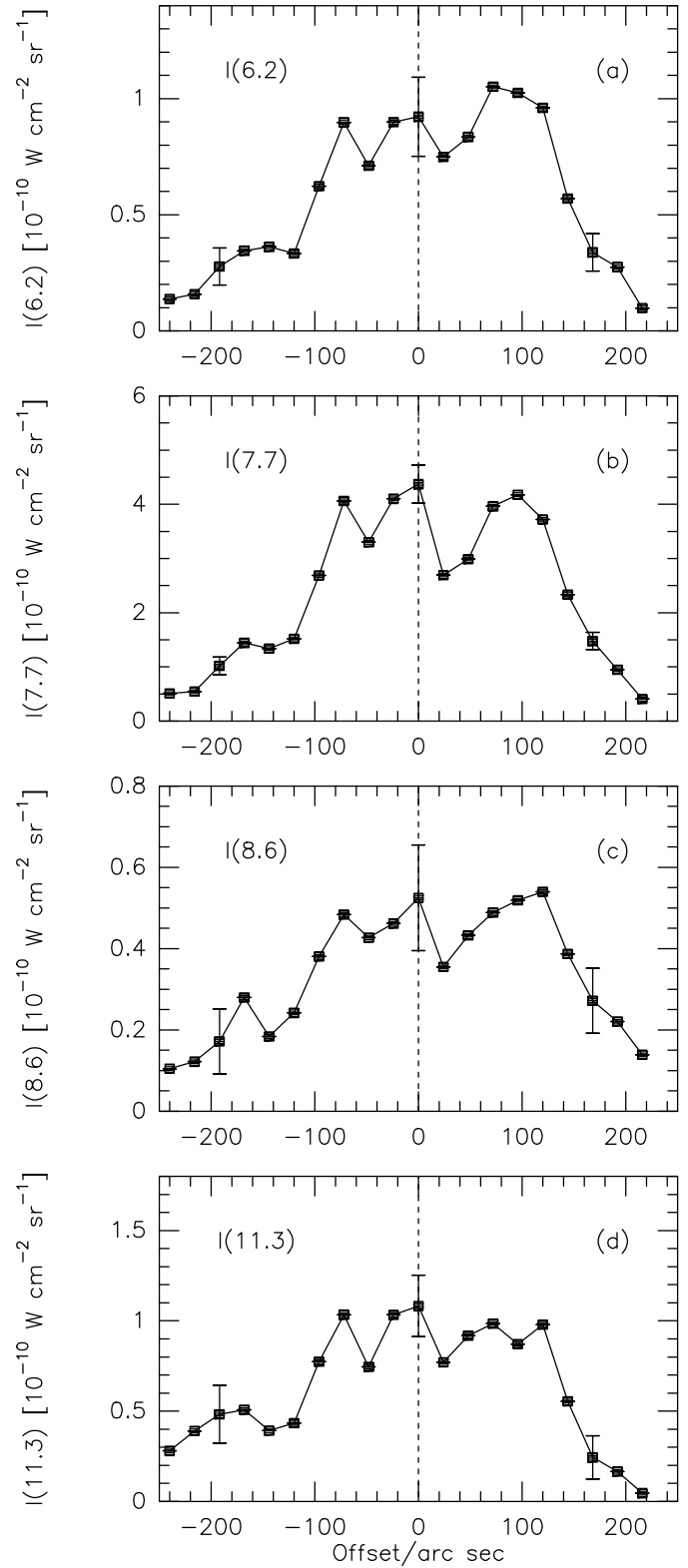}
     \vspace{-0cm}
     \caption{{\bf a-d.} Distribution of the UIR band intensities along the major 
axis of NGC~891. X-axis is the offset from the centre of the galaxy. Representative
{\em statistical} error bars for three positions are shown. The relative point-to-point
calibaration errors are $\sim$10\% and the absolute overall
calibration accuracy is $\sim$30\%} 
     \label{spectra}
\end{figure}

\epsfxsize=12cm
\begin{figure*} [htb]
     \epsfbox{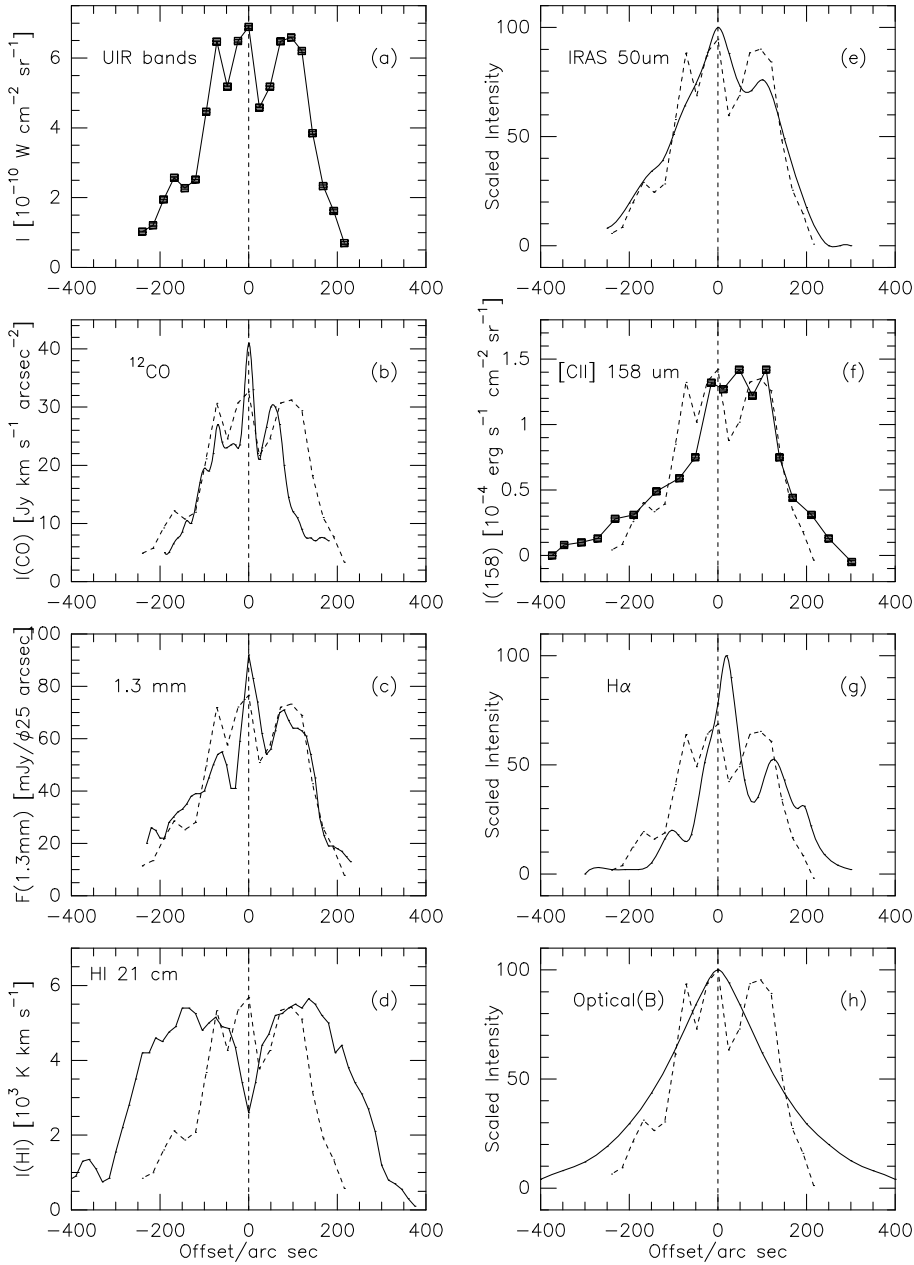}
     \caption{{\bf a-h.} Distributions of different ISM components
along the major axis of NGC~891. X-axis is the offset from the centre of 
the galaxy. The sum of the four UIR band intensities (6.2, 7.7, 8.6, and 
11.3 $\mu$m) is shown in panel {\bf (a)} and, arbitrarily scaled, 
as dotted line in panels {\bf (b) -- (h)}}
     \label{ism}
\end{figure*}

(7) We have fitted the spectra with 
a combination of a linear baseline and four Cauchy line profiles

$$
f(\lambda) = 
 {a_0 \over{1 + ({\Large {\frac{2(\lambda - a_1)}{a_2}})^2}}}
$$

The fit for the spectrum at 120''N is shown in Fig. 3. 
The fit parameters, i.e. the line heights, 
central wavelengths, and linewidths (FWHM) 
for the individual positions were first determined with no parameter
values fixed. Inspection of the resulting central wavelengths and 
line widths showed that there were no significant differences
over the disk or between the disk and the nucleus of NGC~891.
Their mean values together with the standard deviations
are as follows: 
\begin{tabular}{rrr}
6.19 & $\pm$0.01 $\mu$m & (0.22 $\pm$0.02 $\mu$m)\\
7.60 & $\pm$0.02 $\mu$m  &(0.62 $\pm$0.02 $\mu$m)\\
8.48 & $\pm$0.02 $\mu$m & (0.30 $\pm$0.02 $\mu$m)\\
11.24& $\pm$0.01 $\mu$m & (0.30 $\pm$0.05 $\mu$m)\\
\end{tabular}

The central velocities are given in the rest frame of NGC~891 ($V_r$ = 
+530 km~s$^{-1}$).
These values agree well with the wavelengths
and widths observed for the diffuse emission of our Galaxy and for
bright galactic nebulae. 
In order to improve the accuracy of line intensity 
estimates we subsequently
determined the heights of the lines keeping the central wavelength
and line width fixed. The line intensities (areas) $I_{\lambda}$
were then determined from \newline 
${\frac{\pi}{2}}$ $\times$ height $\times$ FWHM.
We give the resulting line intensities together with their (statistical) error 
estimates in columns (2) to (5) of Table~1. The line ratios
together with their statistical error estimates are given in columns (6) to (9).
The external statistical errors (standard deviations) derived 
from the values of 
the northern and southern inner disk are $\pm0.01, \pm0.015, \pm 0.05$, and $\pm0.09$
for the four ratios displayed. 
These are smaller or equal to the statistical errors given in Table 1.
The average values of the band ratios 
$I(8.6)/I(7.7)$,  $I(11.3)/I(7.7)$, and $I(8.6)/I(11.3)$ 
are very closely the same for the northern and southern inner disk 
($r = \pm24''$ to $\pm120''$) (see Table 2). 

We have preferred the
Cauchy line profiles to the Gauss ones. With Cauchy profiles no
additional ``plateau emission'' component between 6.5 -- 9 $\mu$m
is needed; this feature is completely explained by the broad Cauchy line wings.
Also, a better overall fit of the observed spectrum 
is obtained in this wavelength region. Boulanger et al. (1998)
have presented a comparison of Lorentz vs. Gauss profile fits to
ISOCAM CVF spectra. They also find much better fits with the Lorentz
profiles, equivalent to the Cauchy profiles but in frequency instead
of wavelength space. See also their discussion of physical arguments 
justifying the use of Lorentz profiles. One should note that a direct
comparison of the Cauchy/Lorentz feature intensities with the Gauss
ones is not possible: the Cauchy/Lorentz band intesities, especially
for the 7.7 $\mu$m feature, are larger since they include the broad
wing emission.

\subsection{PHT-SS spectra (2.5 -- 4.9 $\mu$m)}

The following results can be listed for the short-wavelength spectra:\\
(1) The 3.3 $\mu$m UIR band is not detected, neither at the central
position nor in the disk. An upper limit of 
$\sim 1~10^{-10}$ W~cm$^{-2}\mu$m$^{-1}$~sr$^{-1}$ can be set to the
height of the line.\\
(2) A stellar continuum is present at $\lambda \la 3 \mu$m. 

\begin{table*}
\caption{Line intensities of UIR bands in NGC~891 (columns 2 -- 5).
The unit is
10$^{-10}$ W cm$^{-2}$ sr$^{-1}$.
Columns (6) - (10)
give the band ratios. The position gives the offset in arc sec from the centre of NGC~891
along the major axis. The {\em statistical} errors are given in units of the last digit in each column. The relative point-to-point calibration errors are $\sim$10\% and the absolute overall 
calibration error is $\sim$30\%}
\begin{center}
\begin{tabular}{|r|crcrcrcr|crcrcrcr|}
\hline
Position &I(6.2) & &I(7.7) && I(8.6) && I(11.3) && 
$\frac{I(6.2)}{I(7.7)}$ && $\frac{I(8.6)}{I(7.7)}$ &&
$\frac{I(11.3)}{I(7.7)}$ && $\frac{I(8.6)}{I(11.3)}$ & \\
(1)  & (2) && (3) && (4) && (5) && (6) && (7) && (8) && (9)& \\
\hline
-240  &0.14 &$\pm$2 &0.51&$\pm$3 &0.10&$\pm$2&0.28&$\pm$4&0.27&$\pm$4&0.20&$\pm$4&0.54&$\pm$8&0.37&$\pm$9 \\
-216  &0.16 &     7& 0.54 &    13&0.12&     7&0.39&    16&0.28&14 &0.22 &14 &0.71 &34 &0.31&22 \\
-192  &0.28 &     8& 1.02 &    16&0.17&     8&0.48&    16&0.27& 9 &0.17 & 8 &0.47 &17 &0.36&20 \\
-168  &0.34&      9& 1.44 &    18&0.28&     9&0.51&    16&0.24& 7 &0.19 & 7 &0.35 &12 &0.55&25 \\
-144  &0.36&      9& 1.34 &    18&0.18&     8&0.39&    16&0.27& 8 &0.14 & 6 &0.29 &12 &0.47&28 \\
-120  &0.33&      9& 1.52 &    19&0.24&     9&0.43&    16&0.22& 6 &0.16 & 6 &0.28 &11 &0.56&29 \\
-96   &0.62&     12& 2.69 &    25&0.38&    10&0.77&    18&0.23& 5 &0.14 & 4 &0.28 & 7 &0.49&18 \\
-72   &0.90&     16& 4.06 &    34&0.48&    12&1.03&    19&0.22& 4 &0.12 & 3 &0.25 & 5 &0.47&14 \\
-48   &0.71&     13& 3.30 &    29&0.43&    11&0.74&    18&0.21& 4 &0.13 & 3 &0.22 & 6 &0.57&20 \\
-24   &0.90&     16& 4.10 &    34&0.46&    12&1.03&    19&0.22& 7 &0.11 & 3 &0.25 & 5 &0.45&14 \\
0     &0.92&     17& 4.37 &    35&0.52&    13&1.08&    17&0.21& 4 &0.12 & 3 &0.24 & 4 &0.48&14 \\
24    &0.75&     12& 2.70 &    23&0.36&     9&0.77&    14&0.27& 5 &0.13 & 3 &0.28 & 6 &0.46&14 \\
48    &0.84&     13& 2.99 &    26&0.43&    10&0.92&    15&0.27& 5 &0.14 & 4 &0.30 & 6 &0.47&13 \\
72    &1.05&     15& 3.96 &    31&0.49&    11&0.98&    16&0.26& 4 &0.12 & 3 &0.24 & 4 &0.50&14 \\
96    &1.02&     15& 4.18 &    31&0.52&    11&0.87&    15&0.24& 4 &0.12 & 3 &0.20 & 4 &0.60&16 \\
120   &0.96&     14& 3.72 &    30&0.54&    11&0.98&    16&0.25& 4 &0.14 & 3 &0.26 & 5 &0.55&14 \\
144   &0.57&      2& 2.33 &     5&0.39&     2&0.56&     3&0.24& 1 &0.16 & 1 &0.23 & 1 &0.70&5  \\
168   &0.34&      8& 1.48 &    16&0.27&     8&0.24&    12&0.22& 6 &0.18 & 6 &0.16 & 8 &1.11&64 \\
192   &0.28&      1& 0.95 &     1&0.22&     1&0.17&     1&0.28& 1 &0.23 & 1 &0.17 & 1 &1.33&10 \\
216   &0.10&      3& 0.41 &     6&0.14&     4&0.05&     8&0.24& 8 &0.34 & 11&     &   &    & \\
\hline
\end{tabular}
\end{center}
\label{tab1}
\end{table*}

\begin{table*}
\caption{UIR band ratios as observed in different parts of NGC~891 (present study) 
and other objects (from literature).}
\begin{center}
\begin{tabular}{|l|llll|l|}
\hline
Object  &
$\frac{I(6.2)}{I(7.7)}$ & $\frac{I(8.6)}{I(7.7)}$ &
$\frac{I(11.3)}{I(7.7)}$ & $\frac{I(8.6)}{I(11.3)}$ & Reference \\
 (1)  & (2) & (3) & (4) & (5) &  \\
\hline
NGC~891 \quad Southern end -168'' to -240''&0.263$\pm$10&0.195$\pm$12&0.52$\pm$8&0.40$\pm$5& Present study\\
NGC~891 \quad Inner disk South -24'' to -120''&0.217$\pm$3&0.130$\pm$8&0.26$\pm$3&0.51$\pm$2&Present study\\
NGC~891 \quad Inner disk North +24'' to +120''&0.260$\pm$6&0.132$\pm$5&0.26$\pm$2&0.52$\pm$6& Present study\\
NGC~891 \quad Northern end +144'' to +192'' &0.250$\pm$18&0.191$\pm$20&0.19$\pm$2&1.04$\pm$18&Present study\\
\hline
Milky Way \quad Diffuse radiation               &0.215 &0.159& 0.238 & 0.676& Mattila et al.(1996)   \\
 ($l = -5^{\circ}, -30^{\circ} ; b = 0$)&  &              &        &              &         \\
\hline 
G300.2-16.8 \quad Cirrus            &              &        & 0.33-0.43    & &Lemke et al. (1998)   \\
$\rho$~Oph Bright rim     &0.32          &0.17    & 0.30         & 0.57 &Boulanger et al.(1996)  \\
\hline                      
Twelve galaxy nuclei(starburst)      &      &      &      &0.73$\pm$0.25& Aitken \& Roche(1984)\\
                                     &      &      &      &(0.40 -- 1.44) &                    \\
\hline
Six Compact HII Regions           &0.10-0.50&0.11-1.7&0.09-0.20&0.7-10&Roelfsema et al.(1996) \\
NGC~1333 \quad $r = 1'' - 11''$      &      &      &      &0.65-0.27& Joblin et al.(1996) \\
\hline
\end{tabular}
\end{center}
\label{tab1}
\end{table*}

\section{Discussion}

\subsection{General properties} 

Despite the much lower level of emission, per
unit mass, and the different environment
of interstellar space which is probed by our NGC~891 spectra, the basic 
properties and ratios of the UIR bands are surprisingly similar to the 
objects with a much higher radiation field, 
the planetary and reflection nebulae, and HII regions. Also, less 
surprisingly, they are very similar to the diffuse disk emission of our 
Galaxy.  This demonstrates that the
carriers of the UIR bands are very resistant to different interstellar
environmental and ageing effects. Also, if the UIR bands are caused by
a mixture of different molecule/grain populations their relative
abundances and physical parameters (e.g. ionization)
must be remarkably constant or, alternatively, the spectral
characteristics of the different molecules/grains must be very similar
both in NGC~891 and in our Galaxy. 

The intensities for the two data sets, i.e. the  NGC~891 disk and 
the diffuse emission of our Galaxy (Mattila et al. 1996), are very similar. 
This is, however, more probably a coincidence than the result of the 
similarity of these two galaxies. The width of the ISM 
distribution in NGC~891 (HI, CO, cold dust) has been 
observed to be $\la$ 15'' (Rupen 1991, Scoville et al. 1993, Gu\'elin
et al. 1993) causing the filling factor in our 24'' aperture to be $\la 0.5$. 
In the case of our Galaxy the beam filling factor
is probably close to 1 but the regions selected were
intentionally avoiding the bright portions of the galactic disk emission. 

In order to produce thermal emission at the level observed in the UIR
peaks of our NGC~891 spectra one needs a temperature of $\ge$ 100 K.
Such high {\em equilibrium} temperatures cannot be reached by grains
in the very low ISRF of the diffuse ISM in a galactic disk.
Thus  our observation of mid-IR emission in the UIR bands
excludes an equilibrium emission and points to emission during thermal spikes.

There is a continuum present both at 6 $\mu$m as well as in the 9 - 10
$\mu$m range. This is in agreement with the observations for 
reflection nebulae, the strongly irradiated $\rho$ Oph
rim (see Boulanger et al. 1996) and the cirrus cloud G300.2-16.8 (Lemke et al.
1998). However, the situation in NGC~891 differs from that encountered 
for the diffuse emission of our Galaxy (Mattila et al. 1996) where no
10 $\mu$m continuum was seen. The 10 $\mu$m continuum in NGC~891 is 
most probably
produced by the emission of transiently heated small particles. The
contribution by stellar photospheres or circumstellar shells is
expected to be negligible (see e.g. Boulade et al. 1996).

\epsfxsize=8.8cm
\begin{figure} [ht]
     \vspace{0cm}
     \epsfbox{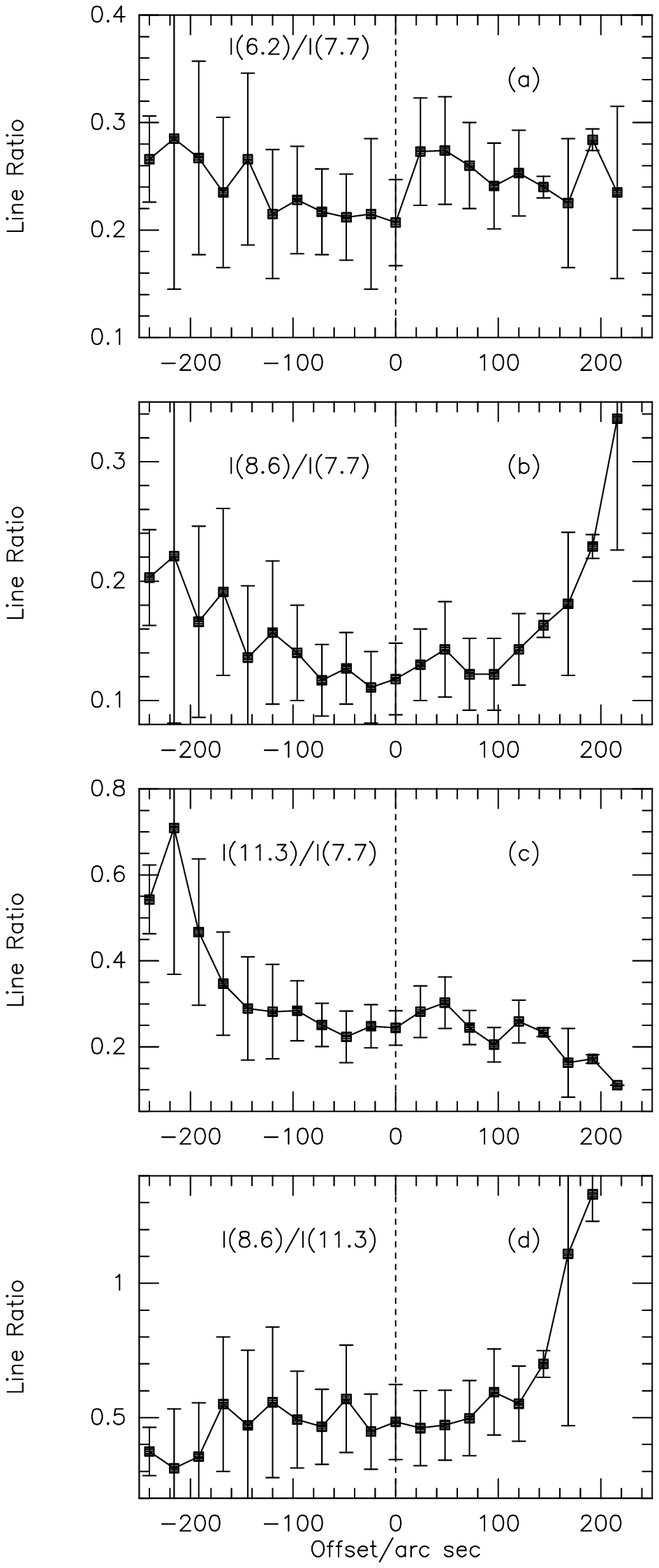}
     \vspace{-0cm}
     \caption{{\bf a-d.} UIR band intensity ratios along the major 
axis of NGC~891. X-axis is the offset from the centre of the galaxy.
Statistical error bars are shown} 
     \label{spectra7}
\end{figure}

\epsfxsize=8.8cm
\begin{figure} [ht]
     \vspace{0cm}
     \epsfbox{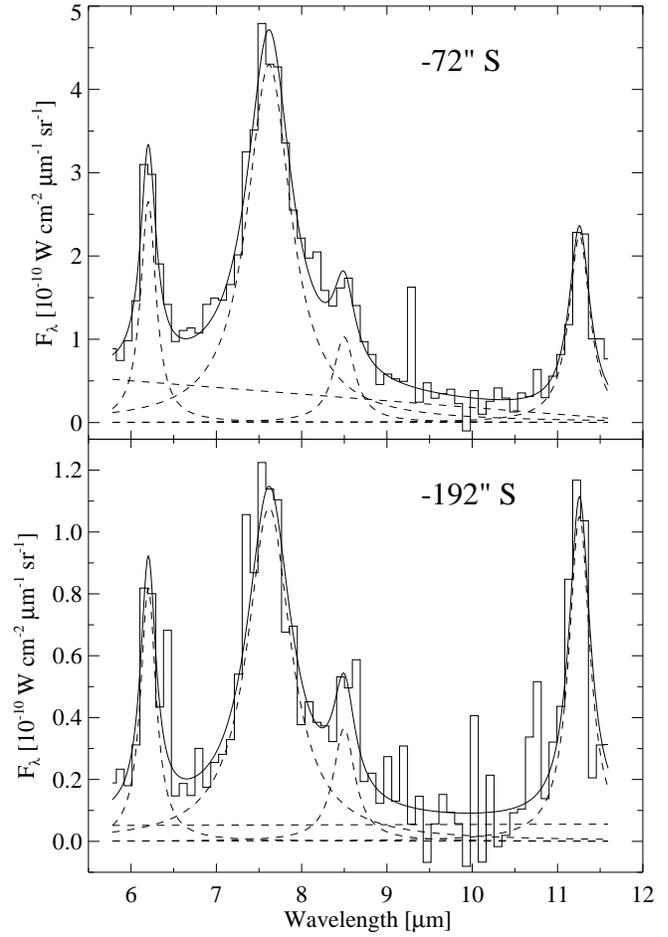}
     \vspace{0cm}
     \caption{ Comparison of the UIR spectra at the -72''~S and -192''~S 
positions. 
Notice the different relative intensities of the 11.3 and 8.6 $\mu$m
bands with respect to the 7.7 $\mu$m band }
     \label{spectra8}
\end{figure}

\epsfxsize=8.8cm
\begin{figure} [ht]
     \vspace{0cm}
     \epsfbox{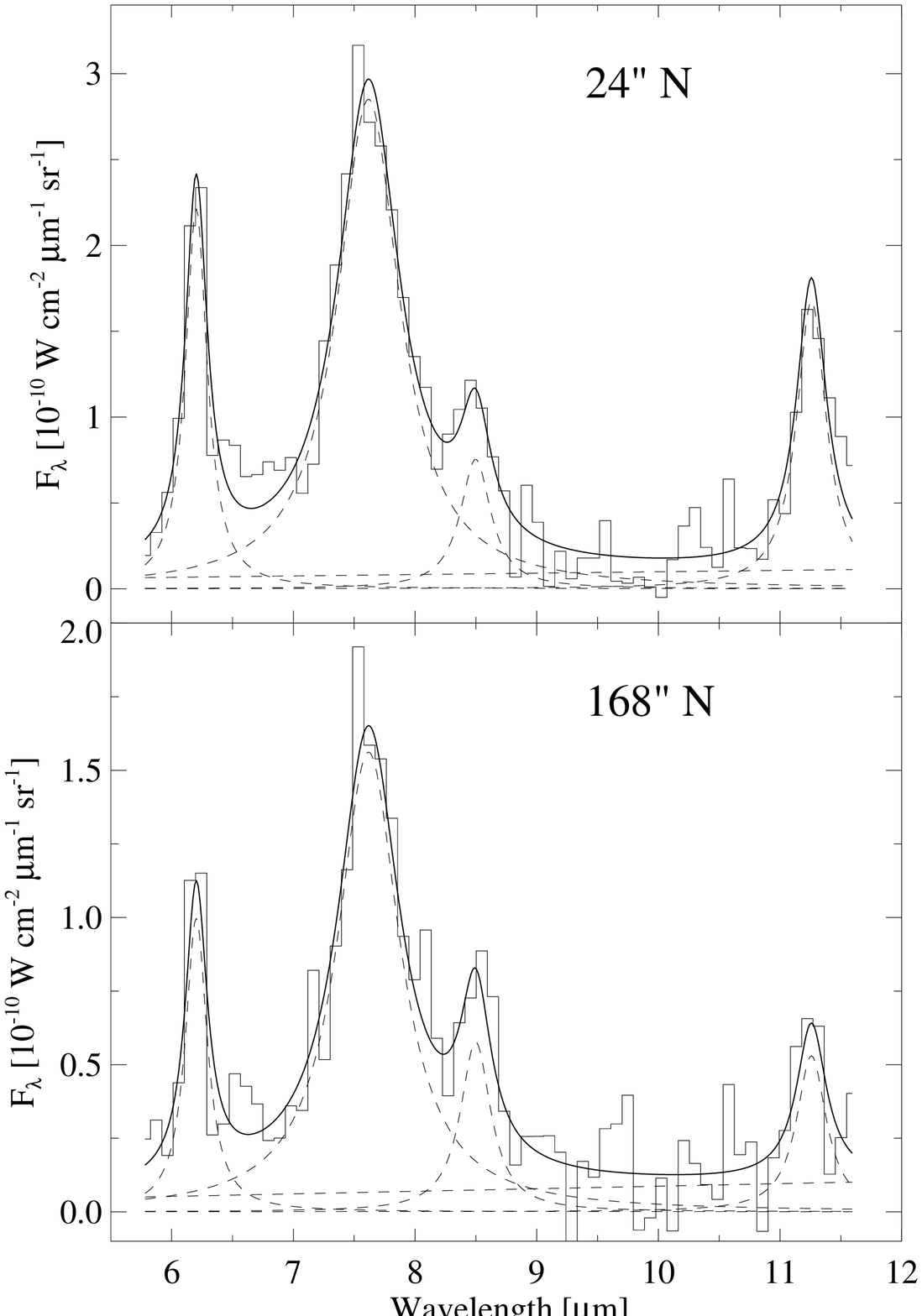}
     \vspace{0cm}
     \caption{Comparison of the UIR spectra at the 24''~N and 168''~N 
positions. 
Notice the different relative intensity of the 8.6 $\mu$m
band with respect to the 11.3 $\mu$m band }
     \label{spectra9}
\end{figure}

\subsection{Radial distributions} 

The line intensities for the four UIR bands along the major axis are shown 
in Fig.~4a -d. Besides
the general trend of decreasing line intensity with increasing distance
from the centre we notice for the four UIR bands the following
characteristic features:\\ 
(1) a central peak;\\ 
(2) a ring of minimum intensity between $\sim$ 20'' to 50'';\\ 
(3) a ring of maximum intensity between $\sim$~70'' to 120'';\\
(4) The profiles are not symmetric:
the minimum appears deeper and the maximum broader and stronger on
the northern side. 

In Fig.~5 we compare the distribution of the UIR band intensity,
represented by the sum of the 6.2, 7.7, 8.6, and 11.3 $\mu$m band intensities, 
with that of other ISM components: \\
(1) cold dust, traced by the 1.3 mm continuum emission intensity 
(Gu\'elin et al. 1993) \\
(2) the molecular gas which is traced by the $^{12}$CO(2-1) integrated line
intensity (Scoville et al. 1993); \\
(3) the atomic hydrogen, traced by the 21 cm integrated line intensity 
(Sancisi \& Allen 1979);  \\
(4) warm dust, traced by the 50 $\mu$m IRAS CPC emission (Wainscoat et al.
1987);\\
(5) ionized carbon, traced by the 158 $\mu$m integrated line intensity
(Madden et al. 1995), and\\
(6) ionized hydrogen, traced by the H$\alpha$ emission (Dettmar 1990,
Dahlem et al. 1994).

It is striking how closely the UIR band intensity follows the $^{12}$CO(2-1)
line and the 1.3 mm continuum dust emission. 
These two data sets have closely the same resolution ($\sim$~24'' - 30'') as our
PHT-S data. Considering the coarser resolution (70'' - 80'') of the
IRAS 50 $\mu$m data the agreement with the UIR band distribution is also 
good. The HI 21 cm distribution, on the other hand, is distinctly different.
It has a depression in the central area, which suggests a hole in the
HI distribution (but can be partly caused by absorption of the 
continuum emission from the galactic nucleus); the maximum 21 cm intensity 
occurs further out, 
and the radial extent is larger than that of the UIR band emission.  
Details of the [CII] distribution
have been smoothed out by the larger beam ($\sim$~1') used
in these observations but it is nevertheless seen that the [CII]  
distribution has a smaller radial extent than the UIR 
bands along the southern major axis.

The UIR band intensities depend on two factors, the abundance of
their carriers and the ambient radiation density of the exciting photons.
It is usually assumed that UV photons below $\lambda < 4000 \ang$ are
needed to excite the UIR band carriers (See, however, Uchida et al. 1998
for other evidence). We show in Fig.~5h the distribution
of the optical (3400 - 6900 $\ang$) emission of the van der Kruit \& Searle 
(1981) disk model for NGC~891. By comparing the different distributions
in Figs.~5a-h we come to the following conclusions:\\
(1)the UIR band carriers have a distribution very similar to CO and
large cold dust grains;\\
(2) a large-scale association of the UIR band emission with neutral hydrogen is not
present; \\
(3) the UIR band emission does not follow the ionized gas distribution 
(H$\alpha$, [CII] 158 $\mu$m); thus an essential part of the UIR band emission 
must come from the non-ionized ISM component.

\epsfxsize=8.8cm
\begin{figure} [ht]
     \vspace{0cm}
     \epsfbox{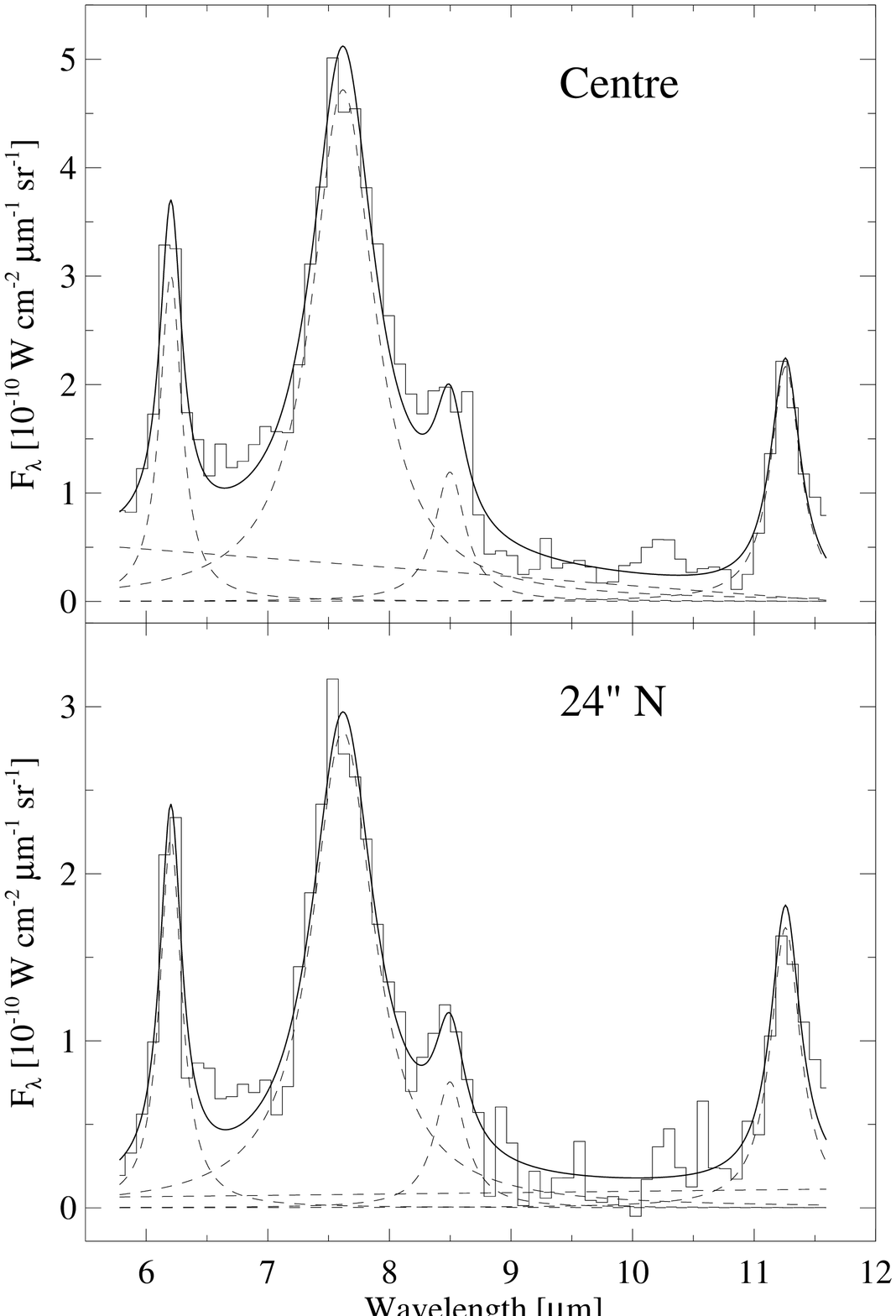}
     \vspace{0cm}
     \caption{Comparison of the UIR spectra at the centre and 24'' N positions. 
Notice the different relative intensity of the 6.2 $\mu$m
band with respect to the 7.7 $\mu$m band }
     \label{spectra10}
\end{figure}

\subsection{Variations of the band ratios} 

The band ratios  $I(6.2)/I(7.7)$, $I(8.6)/I(7.7)$, $I(11.3)/I(7.7)$, and
$I(8.6)/I(11.3)$ 
as observed along the major axis of NGC~891 are given in Table 1, and are
plotted in Fig.~6.

By inspecting Fig.~6 and Table 1 we notice that
there are some significant trends and differences of the line ratios 
between the different parts of NGC~891:\\
(1) On the southern axis, outside of the molecular ring the 
11.3/7.7 $\mu m$ ratio increases with
increasing offset, from $\sim$ 0.23 at r $\sim$ -100'' to $\sim$ 0.50
at r $\sim$ -200''. This difference
is most clearly demonstrated by the two spectra shown in Fig.~7.
The 8.6/7.7 $\mu$m ratio shows the same tendency,
increasing from $\sim$ 0.14 at r $\sim$ -100'' to $\sim$ 0.20
at r $\sim$ -200''. The 8.6/11.3 $\mu$m
ratio decreases slightly over this range.\\
(2) On the northern axis, outside the molecular ring, the 8.6/7.7 $\mu$m ratio
increases from $\sim$ 0.15 at r $\sim$ 100'' to $\sim$ 0.30
at r $\sim$ 200''. Over the same range there is a modest decrease in
the 11.3/7.7 $\mu$m ratio. As a result the 8.6/11.3 $\mu$m ratio shows
even a stronger increase than the 8.6/7.7 $\mu$m ratio: 
from $\sim$ 0.6 at r $\sim$ 100'' to $\sim$ 1.2 at r $\sim$ 200'' 
(see also Fig.~8).\\
(3) Although the 6.2 and 7.7 $\mu$m intensity distributions are
qualitatively quite similar (see Fig.~4) 
the ratio 6.2/7.7 $\mu$m shows a rather
peculiar behaviour: its value is constant, $\sim$ 0.25, over the whole
northern as well as the outer part of the southern major axis but
between $r \sim$ 0'' and -120'' it drops to $\sim$ 0.22. (see also
Fig.~9). This cannot
be explained by calibration differences between revolutions 656 and 788 as the positions at 0'' and +120'' which were observed during both revolutions
gave consistent results.\\
(4) The line ratios for the central position
do not differ from the values of the inner galactic disk.

For comparison with the different areas of NGC~891 we show in Table~2 
the line ratios as given in literature
for a number of different galactic and extragalactic objects.
The values for two positions
in the inner Milky Way disk have been observed with the same instrument
(PHT-S) and applying similar analysis methods (e.g. Cauchy fits) as
for NGC~891. The Milky Way values are seen to agree with the NGC~891
inner disk values except for the $I(8.6)/I(11.3)$ ratio which 
is similar to the northern major axis values.
The $I(11.3)/I(7.7)$ value for the G300.2-16.8 cirrus cloud, derived from
ISOPHOT filter photometry, is in reasonably good agreement with
the NGC~891 southern major axis range of values. The band ratios
for the $\rho$ Oph Bright rim were derived from ISOCAM CVF observations
(Boulanger et al. 1996) using a similar fitting procedure (Lorentz profiles)
as for NGC~891. The band ratios are in good agreement with the NGC~891
inner disk values except, perhaps, for the $I(6.2)/I(7.7)$ ratio which
appears to be somewhat larger. The range of values in $I(8.6)/I(11.3)$
covered by the sample of twelve galaxy nuclei (Aitken \& Roche 1984)
brackets the NGC~891 inner disk values; the average value of the band ratio 
for the galaxy nuclei, 0.73$\pm0.25$, is larger than the value for the
NGC~891 inner disk and centre.

\subsection{Band ratios in the PAH model}

   
There are several possible
mechanisms which can influence the PAH band ratios and can potentially
explain their variations as observed in NGC~891:


(1) {\em The dehydrogenation of PAHs.}
The strength of the C--C vibration bands (6.2 and 7.7 $\mu$m) depends
only on the PAH skeleton structure, whereas the C--H band strengths
(8.6 and 11.3 $\mu$m) depend on the number of H atoms in the
molecule. The degree of dehydrogenation is measured by 
$f_{\rm H}$ which is the ratio between the numbers of actual and possible 
H atoms in the PAH molecule. Thus $f_{\rm H}$ = 1 for complete 
hydrogen coverage and $f_{\rm H}$ = 0 for fully dehydrogenated molecule.
The 8.6 $\mu$m band strength is proportional to the number
of H atoms. For the 11.3 $\mu$m band the dependence is somewhat more
complicated: with increasing dehydrogenation its strength first slightly
increases until $f_{\rm H} \approx$ 0.7 and then monotonically drops for
$f_{\rm H} < 0.7$. This is caused by the duo and trio H bonds, see
Schutte et al. 1993. For strong dehydrogenation it is thus possible
to obtain a small intensity ratio of the 11.3 $\mu$m to the 6.2 and 7.7 $\mu$m
bands, corresponding to the ISM observations
(see L\'eger et al. 1989, Jourdain de Muizon et al. 1990, Schutte et al. 1993).
The dehydrogenation is thought to occur through UV photodissociation, thus
becoming monotonically weaker when the ISRF weakens.

This mechanism fails, however, to explain why the band ratios are so
similar in high- and low-ISRF objects although the dehydrogenation degrees
are expected to be very different. Furthermore, it does not explain
why also the 8.6/11.3 $\mu$m ratio is much higher in the ISM than 
predicted by the dehydrogenated PAH model.

(2) {\em The fraction of PAH cations, [PAH$^+$]/[PAH].}
A probable solution to the band ratio problem has emerged from recent
laboratory and theoretical results for PAH cations (for a review see
Allamandola et al. 1995): the 6.2, 7.7, and 8.6 $\mu$m band cross sections
for PAH$^+$s are typically a factor of 10 larger than for neutral PAHs
whereas the 11.3 $\mu$m band is much less influenced.
The ISM 11.3/7.7 $\mu$m ratio could thus be understood if a substantial
fraction of the PAHs are ionized.
The ionization degree
results from an equilibrium between ionizing events, depending on the
intensity of the ISRF, and recombination events, depending on the local
electron density. In regions with high UV radiation density and low electron
density, such as reflection nebulae, the fraction of ionized PAHs approaches
unity (Omont 1986). In low-ISRF diffuse medium the situation is less clear. 
Verstraete
et al. (1990) have estimated that for the two compact PAHs, coronene and
pyrenene and for a solar-neighbourhood ISRF the ionization degree
([PAH$^+$]/[PAH]) is $\sim$6\%. For the same PAH molecules Bakes \& Tielens
(1994) and Salama et al. (1996) have calculated similarly low [PAH$^+$]/[PAH]
ratios in moderately dense ($n_{\rm H} \ge$ 200 cm$^{-3}$) diffuse clouds.
Joblin et al. (1996) have found that the 8.6/11.3 $\mu$m ratio in the
reflection nebula NGC~1333 decreases by a factor of $\sim$2 between the
position of the star and outer nebula in accordance with the calculated degree
of PAH ionization.

(3) {\em The fraction of compact and non-compact species in the PAH ion 
mixture.}
Absorption spectra of a number of compact and non-compact PAH cations were 
measured by Hudgins et al. (1994) and Hudgins \& Allamandola (1995). Roelfsema
et al. (1996) inspected these results especially in view of the large variation
of the 8.6/7.7 $\mu$m band ratio they observed in compact HII regions (see Table~2).
They found
that the addition of less condensed PAHs to the mixture
enhances the 8.6 $\mu$m region relative to the 7.7 $\mu$m region.

\subsection{Interpretation of the observed band ratios}

We will now make an attempt to interpret the observed band ratios and
their variations in NGC~891 in terms of the PAH model.

(1) With reference to point (1) in Sect. 4.3 we notice that the UIR band 
emission, 
especially in
the 8.6 and 11.3 $\mu$m bands, is more
extended on the southern than on the northern side. Also the 1.3 mm cold dust 
emission shows the same kind of 
southern extension with excess emission at $\sim$ -200''. The $^{12}$CO
and the H~I 21-cm emission do not have any excess in this
area, but there is a pronounced 21-cm excess beyond $\sim$ -300''.
The H$\alpha$ emission (corrected for extinction), on the other hand, 
shows a distinct deficiency in this area.  Since
there is no deficiency of atomic or molecular hydrogen in this area 
the deficiency of ionized hydrogen must
be caused by a smaller number of hot young stars. 
Thus the UV ISRF is weaker in this area
as compared to the corresponding regions at the northern major axis.
We interpret the simultaneous increase of the 11.3/7.7 $\mu$m and 
8.6/7.7 $\mu$m ratios to be caused 
mainly by increased hydregenation of PAHs in this region with low UV ISRF.

(2) The behaviour of the ratios 8.6/7.7 $\mu$m, 11.3/7.7 $\mu$m, and
8.6/11.3 $\mu$m towards the end of the northern major axis (point 2
in Sect. 4.3) calls for a different explanation. An increasing fraction
of PAH cations relative to neutral PAHs can explain the 
increase of the 7.7/11.3 $\mu$m and 8.6/11.3 $\mu$m ratios. However,
the increase of the 8.6/7.7 $\mu$m ratio can not be explained 
this way. In view of the results of Roelfsema et al. (1996) for
galactic HII regions (see Sect. 4.4, point (4) above) we suggest that this 
effect is due to 
the increasing contribution of non-compact PAH cations, residing preferentially
in high-luminosity HII regions. This interpretation is in accordance with
the observed higher H${\alpha}$ luminosity indicating higher ionization at
the northern part of the major axis.

(3) The modest depression of the 6.2/7.7 $\mu$m ratio on the inner
southern major axis between 0'' and $\sim$ -100'' is more difficult
to explain. It could be due to a lower average temperature of PAHs
during their temperature spikes. This could happen if the mean energy of the
UV photons were lower or the PAH size larger in this region. It is difficult
to judge on the basis of the available data whether either of these
prerequisites is fulfilled.

(4) Assuming that the width of the 3.3 $\mu$m UIR band in NGC~891 
is the same as in the diffuse emission of our Galaxy, 
$\Delta \lambda \approx 0.13 \mu$m (Tanaka et al. 1996),
the upper limit of the 3.3 $\mu$m band intensity in the NGC~891 disk can be
estimated to be 
$\sim 2~10^{-11}$ W~cm$^{-2}$~sr$^{-1}$. The 
11.3 $\mu$m band intensity is $\sim 6~10^{-11}$ W~cm$^{-2}$~sr$^{-1}$ 
(see Table~1), which gives a lower limit of $\sim$3 for 
the 11.3/3.3 $\mu$m ratio.
Following the analysis of Jourdain de Muizon et al. (1990) this allows
us to set for the average PAH emission temperature an upper limit of
$\langle T \rangle \la 640$~K which leads to an estimate for the lower
limit of the average PAH size (number of atoms) of $N_{\rm tot} \ga 70$.
These limits indicate that the temperature is at least somewhat lower 
and the PAH 
size at least somewhat larger
than the typical values found for HII regions and reflection nebulae
(see L\'eger et al. 1989, Jourdain de Muizon et al. 1990). 
 
\section{Conclusions}  

(1) We have observed for the first time the UIR band spectra at 6.2, 7.7, 8.6,
and 11.3 $\mu$m in the diffuse emission of an external galaxy disk, the
nearby edge-on Sb spiral NGC~891.
 
(2) The UIR band emission between 5.9 and 11.7 $\mu$m is $\sim$9\% 
of the total infrared emission of the galaxy.

(3) The UIR band emission is visible along the major axis of the galaxy up to a
distance of $\sim$10 kpc. 

(4) The distribution of the UIR band intensities 
is closely similar to that of the 1.3 mm continuum emission by cold dust and
to that of the $^{12}$CO(1-0) line intensity, whereas the 21-cm HI emission
has a substantially broader distribution. The distributions of the 
H${\alpha}$ and [CII] 158 $\mu$m
line emissions are also different from the ones for the UIR bands. 
We conclude that the UIR band carriers are mainly associated with 
areas of neutral molecular gas.   

(5) The UIR band ratios show substantial variations along the major axis of
NGC~891. Adopting the PAH model the band ratios suggest 
a low dehydrogenation
degree of the PAHs at the southern end of the major axis,
and 
an increase of the fraction of PAH cations, especially the
non-compact  ones, along the northern major axis of NGC~891.

\small
{\em Acknowledgements.} We gratefully acknowledge the 
financial support for this research project by the University of Helsinki
Research Fund and the Academy of Finland (Project 1011055). 

{}

\end{document}